\documentclass{aa}  

\usepackage{graphicx}
\usepackage{color}
\usepackage{arydshln}
\usepackage{natbib}
\usepackage{savefnmark}
\bibpunct{(}{)}{;}{a}{}{,}%to follow the A&A style
\usepackage{txfonts}
\begin{document} 

\def\sun{\hbox{$\odot$}}
\def\degr{\hbox{$^\circ$}}
\def\arcmin{\hbox{$^\prime$}}
\def\arcsec{\hbox{$^{\prime\prime}$}}

   \title{Sub-arcsecond imaging of the water emission in Arp~220\thanks{Based on observations carried in ALMA programs  
     ADS/JAO.ALMA\#2011.0.00018.SV and ADS/JAO.ALMA\#2012.1.00453.S, and with the IRAM 30~m telescope under project numbers 189-12 and 186-13.} 
     \thanks{We dedicate this work to the memory of Fred Lo.}}

   \author{S. K\"onig
          \inst{1}
          \and
          S. Mart\'{i}n
	  \inst{2,3}
          \and
          S. Muller
	   \inst{1}
         \and
          J. Cernicharo
          \inst{4}
         \and
          K. Sakamoto
          \inst{5}
	  \and
          L.~K. Zschaechner 
	   \inst{6}
         \and
	  E.~M.~L. Humphreys
   	  \inst{7}
         \and
	  T. Mroczkowski
   	  \inst{7}
	 \and
          M. Krips
          \inst{8}
	 \and
          M. Galametz
          \inst{9,7}
	 \and
          S. Aalto
          \inst{1}
	 \and
	  W.~H.~T. Vlemmings
	  \inst{1}
	  \and
          J. Ott
	   \inst{10}
	  \and
          D.~S. Meier
	   \inst{11}
	  \and
          A. Fuente
	   \inst{12}
	  \and
          S. Garc\'{i}a-Burillo
	   \inst{13}
	  \and
	  R. Neri
	   \inst{8}
          }

   \institute{Chalmers University of Technology, Department of Earth and Space Sciences, Onsala Space Observatory, 43992 
	     Onsala, Sweden %\\
              \email{sabine.koenig@chalmers.se}
         \and
             European Southern Observatory (ESO), Alonso de C\'{o}rdova 3107, Vitacura, Casilla 19001, 763 0355, Santiago, Chile %\\
         \and
             Joint ALMA Observatory, Alonso de C\'{o}rdova 3107, Vitacura, Casilla 19001, 763 0355, Santiago, Chile %\\
	 \and 
	     Grupo de Astrofísica Molecular, Instituto de CC. de Materiales de Madrid (ICMM-CSIC), Sor Juana In\'{e}s de la Cruz 3, 
             Cantoblanco, 28049, Madrid, Spain %\\
	 \and 
	     Institute of Astronomy and Astrophysics, Academia Sinica, PO Box 23-141, 10617, Taipei, Taiwan %\\
	 \and 
	     Max Planck Institute for Astronomy, K\"onigstuhl 17, 69117 Heidelberg, Germany %\\
	 \and 
	     European Southern Observatory (ESO), Karl-Schwarzschild-Str. 2, 85748, Garching bei M\"unchen, Germany %\\
	 \and 
             Institut de Radioastronomie Millim\'etrique (IRAM), 300 rue de la Piscine, Domaine Universitaire, 38406, Saint 
  	     Martin d'H\`eres, France%\\
	 \and 
	     Laboratoire AIM, CEA/IRFU/Service d'Astrophysique, B\^{a}t. 709, 91191, Gif-sur-Yvette, France %\\
         \and
             National Radio Astronomy Observatory (NRAO), P.O. Box O, 1003 Lopezville Road, Socorro, NM 87801, USA %\\
         \and
             New Mexico Institute of Mining and Technology, Socorro, NM, USA %\\
	 \and
	     Observatorio Astronómico Nacional (OAN, IGN), Apdo 112, 28803, Alcal\'{a} de Henares, Spain %\\
	 \and
	     Observatorio de Madrid, OAN-IGN, Alfonso XII, 3, 28014, Madrid, Spain %\\
             }

   \date{Received ; accepted }

  \abstract
   {}
   {Extragalactic observations of water emission can provide valuable insights into the excitation of the interstellar medium. In particular 
    they allow us to investigate the excitation mechanisms in obscured nuclei, i.e. whether an active galactic nucleus or a starburst dominate.}
   {We use sub-arcsecond resolution observations to tackle the nature of the water emission in Arp~220. ALMA Band~5 science 
    verification observations of the 183~GHz H$_{\rm 2}$O\,3$_{\rm 13}-2_{\rm 20}$ line, in conjunction with new ALMA Band~7 
    H$_{\rm 2}$O\,5$_{\rm 15}-4_{\rm 22}$ data at 325~GHz, and supplementary 22~GHz H$_{\rm 2}$O\,6$_{\rm 16}-5_{\rm 23}$ VLA observations, are used 
    to better constrain the parameter space in the excitation modelling of the water lines.}
   {We detect 183~GHz H$_{\rm 2}$O and 325~GHz water emission towards the two compact nuclei at the center of Arp~220, being brighter in Arp~220 
    West. The emission at these two frequencies is compared to previous single-dish data and does not show evidence of variability. The 183 and 
    325~GHz lines show similar spectra and kinematics, but the 22~GHz profile is significantly different in both nuclei due to a blend with an 
    NH$_{\rm 3}$ absorption line.}
   {Our findings suggest that the most likely scenario to cause the observed water emission in Arp~220 is a large number of independent masers 
    originating from numerous star-forming regions.}


\titlerunning{Sub-arcsecond water emission in Arp~220}

   \maketitle

%
%________________________________________________________________

\section{Introduction} \label{sec:intro}

At a luminosity distance of 78~Mpc (1\arcsec\,=\,378~pc), \object{Arp~220} makes for one of the most interesting objects of study in the nearby 
Universe. It is a suitable target to examine a large number of different tracers of the interstellar medium (ISM) sampling different physical 
conditions on a large range of spatial scales. Its high infrared luminosity 
\citep[L$_{\rm IR}$\,=\,1.4\,$\times$\,10$^{\rm 12}$~L$_{\sun}$,][]{soi87} makes Arp~220 a good local proxy for high-redshift (ultra-)luminous 
infrared galaxies ((U)LIRGs). As a result, we now know that Arp~220 is the result of a merger \citep{arp66,nil73,sak99} -- the two remnant nuclei, 
Arp~220 East and Arp~220 West, are separated by only $\sim$380~pc at the center of Arp~220 \citep[e.g.,][]{dow07,aal09,sak09,mar11}. They are each 
embedded in their own rotating gas disks in the foreground of a kpc-scale molecular gas disk \citep[e.g.,][]{sco97,sak99,koenig12}. The different 
origins of the two nuclei manifest themselves as a misalignment in the rotation axes of the gaseous disks. Whether AGN or the powerful starburst 
associated with the central activity, or a mixture of the two, facilitates the bright appearance of this prototypical ULIRG at many wavelengths is 
still under debate \citep[e.g.,][]{smi98,lon06,dow07,sak08}.\\
\indent
The physical conditions in Arp~220 have been probed using a number of tracers at radio, mm and sub-mm wavelengths 
\citep[e.g.,][]{sco97,sak99,sak08,sak09,aal07,aal09,aal15,dow07,mar11,mar16,gon12,koenig12,wil14,ala15,ran15,sco15,tun15,mar16,var16}. Among others, 
the water lines in the radio and mm wavelength regimes have been targeted \citep[e.g.,][]{cer06,gal16,zsch16}. Water emission is an excellent tracer 
of physical conditions, such as temperature and density of the molecular gas, in the highly obscured innermost parts of external galaxies. The water 
lines at 22, 183, 321 and 325~GHz rest frequency have been observed as masers in various galactic sources \citep[e.g., in star-forming regions 
and circumstellar envelopes around evolved stars,][and references therein]{cer90,cer94,gon95,cer96,cer99,gon98,gon99,deb05,lef11,bar12,ric14}. In 
extragalactic sources, the 22~GHz line is most commonly used to search for maser emission -- more than 150 sources have been found to emit water 
maser emission, mostly in galaxies with active galactic nuclei \citep[AGN, e.g.,][]{lo05,pes16}. The 183~GHz line has been detected towards 
\object{NGC~3079} \citep{hum05}, Arp~220 \citep{cer06,gal16}, and most recently towards \object{NGC~4945} \citep{hum16}. In both NGC~3079 and 
NGC~4945, the 183~GHz maser emission is associated with the AGN. This seems also true for the water maser emission at 321~GHz that was recently 
reported towards \object{Circinus} and NGC~4945 \citep{hag13,hag16,pes16}. All these are classified as megamasers, i.e. the isotropic luminosity is 
higher by a factor of more than 10$^{\rm 6}$ times than what is found for typical emission of Galactic water masers 
\citep[at 22~GHz: $\sim$10$^{\rm -4}$~L$_{\sun}$, e.g.,][]{lo05}. Recently, the 22~GHz H$_{\rm 2}$O\,6$_{\rm 15}-5_{\rm 23}$ has been tentatively 
detected in both nuclei in Arp~220 \citep{zsch16}. At 183~GHz, single-dish observations yielded convincing evidence for the presence of emission of 
the H$_{\rm 2}$O\,3$_{\rm 13}-2_{\rm 20}$ line in emission \citep{cer06,gal16}.\\
\indent
\citet{cer06} presented a first analysis of the 183~GHz water emission in Arp~220 and showed that a combination of the three major water lines at 
22, 183 and 325~GHz is necessary to properly model the excitation in the galaxy center. In this paper, we combine subarcsecond-scale ALMA 
observations of H$_{\rm 2}$O 22 and 325~GHz with the 183 GHz line resolved for the first time with the newly installed ALMA Band~5 receivers 
\citep{bel09,bil10} to refine the modelling of the excitation conditions in Arp~220.\\
\indent
In Sect.\,\ref{sec:obs} the observations, data reduction and analysis are described, in Sect.\,\ref{sec:results} we present the results, and in 
Sect.\,\ref{sec:discussion} we discuss their implications.

%__________________________________________________________________

\section{Observations} \label{sec:obs}

\begin{figure*}[ht]
  \begin{minipage}[hbt]{0.33\textwidth}
  \centering
    \includegraphics[width=0.88\textwidth]{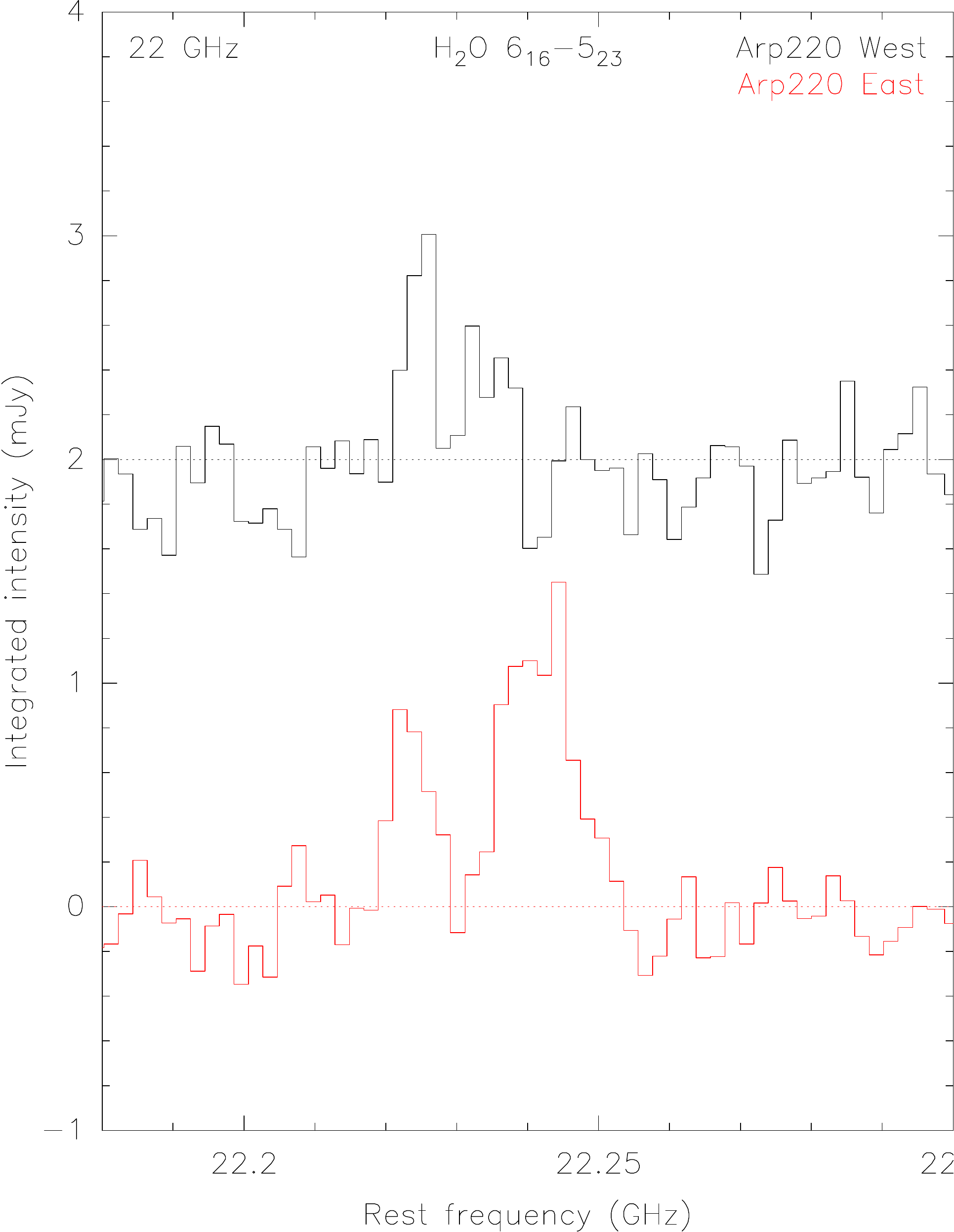}
  \end{minipage}
  \begin{minipage}[hbt]{0.33\textwidth}
  \centering
    \includegraphics[width=0.9\textwidth]{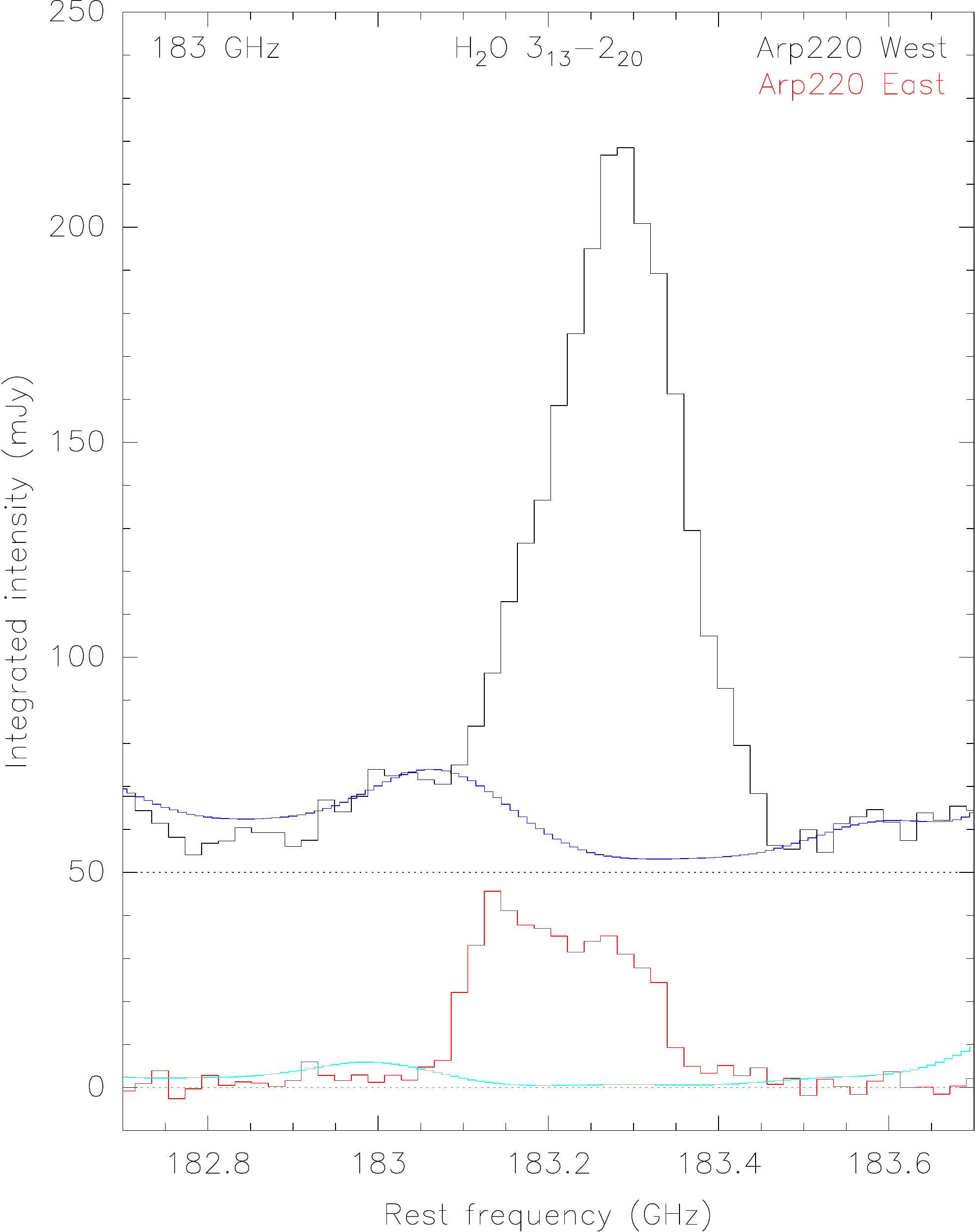}
  \end{minipage}
  \begin{minipage}[hbt]{0.33\textwidth}
  \centering
    \includegraphics[width=0.9\textwidth]{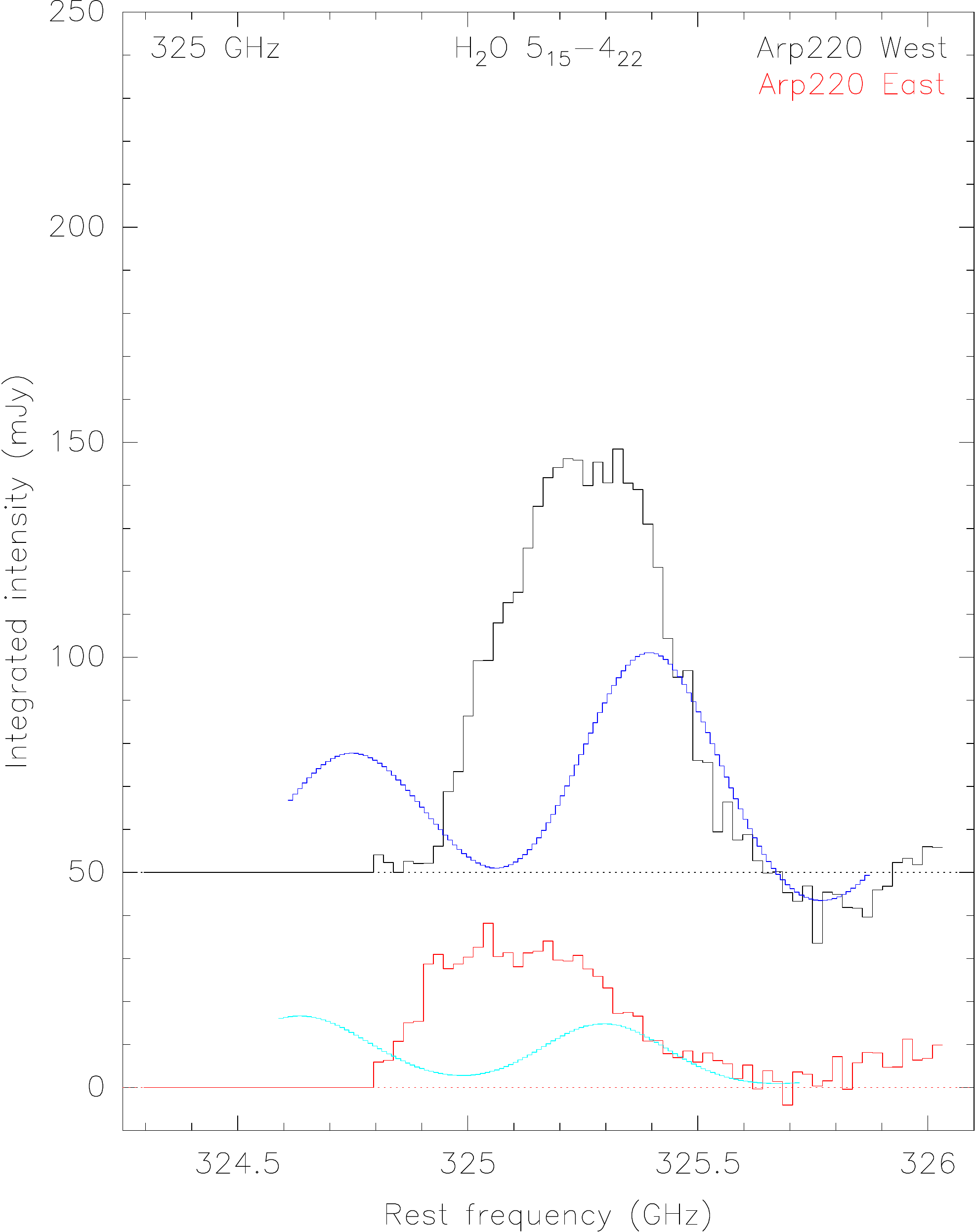}
  \end{minipage}
  \caption{\footnotesize Water emission from Arp~220 West (black) and Arp~220 East (red) at 22 (\textit{left}), 183 (\textit{center}) and 
   325~GHz (\textit{right}). The spectra were extracted from data convolved with the same beam of 0.79\arcsec\,$\times$\,0.71\arcsec. An 
   artificial offset in intensity was introduced to better show the spectra for each nucleus separately. The dotted lines show the zero-
   intensity levels. The 22~GHz spectrum has been corrected for contamination by the NH$_{\rm 3}$\,(3,1) absorption line. The dark and 
   light blue curves in the 183 and 325~GHz spectra represent the synthetic spectrum used to evaluate the contamination of the water line by other 
   emission lines (Mart\'{i}n et al. in prep.).}
  \label{fig:spectra}
\end{figure*}

\subsection{ALMA}

\subsubsection{Band~5} \label{subsubsec:obs_band5}

ALMA observations of the H$_{\rm 2}$O\,3$_{\rm 13}-2_{\rm 20}$ line at 183.310~GHz rest frequency in \object{Arp~220} were obtained on 2016 
July 16th, as part of the Band~5 science verification observations. The dual polarisation Band~5 receivers, designed and developed by the 
Group for Advanced Receiver Development at Chalmers University and Onsala Space Obervatory (GARD, Sweden) and the STFC Rutherford Appleton 
Laboratory (UK), cover the frequency range between 163 and 211~GHz. The receivers provide separated sidebands and an instantaneous bandwidth 
coverage of 8~GHz \citep{bel09,bil10,bil12}. The array was composed of 12 antennas equipped with Band 5 receivers, in a configuration with 
baselines ranging from 30~m to 480~m. The weather conditions were excellent -- about 0.3~mm precipitable water vapor. One spectral window, with a 
bandwidth of 1.875~GHz and spectral resolution of 976.562 kHz, was centered on the water line. Additional spectral windows were placed on the 
HNC\,2$-$1, CS\,4$-$3, and CH$_{\rm 3}$OH\,4$_{\rm 31}-3_{\rm 30}$ transitions, with bandwidths of 1.875~GHz and 0.9375~GHz, at spectral 
resolutions of 976.562~kHz and 488.281~kHz, respectively. For analysis purposes, the resolution was averaged to 20~km\,s$^{\rm -1}$ channels. The 
bandpass response of the antennas was calibrated from observations of the bright quasar \object{J1924-2914}, with \object{J1516+1932} being used 
for primary gain calibration, and \object{J1550+0527} for flux calibration. The continuum emission is strong enough for self-calibration of the 
data. However, it was difficult to determine line-free channels to derive the continuum contribution, especially for Arp~220 West. A model of the 
source was constructed after a shallow clean, using line-free channels in spectral windows 2 and 3 (upper sideband). Then, gain phase solutions 
were determined for each integration and applied to all spectral windows, including the lower sideband. The calibrated visibilities were 
deconvolved using the ``tclean'' task in CASA 4.7.0\footnote{http://casa.nrao.edu/} \citep{mcm07} resulting in a beam size of 
0.79\arcsec\,$\times$\,0.71\arcsec ($\sim$0.30~kpc\,$\times$\,0.27~kpc), and individual data cubes were created for each spectral window. 
The resulting sensitivities are shown in Table\,\ref{tab:water_properties}. After calibration and imaging within CASA, the visibilities were 
converted into FITS format and imported in the GILDAS/MAPPING\footnote{http://www.iram.fr/IRAMFR/GILDAS} for further analysis.

\subsubsection{Band~7} \label{subsubsec:obs_band7}

The Band~7 ALMA data covering the H$_{\rm 2}$O\,5$_{\rm 15}-4_{\rm 22}$ line at 325.153~GHz rest frequency were obtained on June 16 and July 17, 
2014 as part of a full spectral line survey of ALMA Bands~6 and 7 (Mart\'{i}n et al., in prep.; Project 2012.1.00453.S). The array was composed of 
39 antennas. The data were self-calibrated and smoothed to channel widths of 10MHz. A more detailed description of 
the calibration of the data will be presented in Mart\'{i}n et al. (in prep.). For analysis purposes, the channels were further binned to 
widths of 20~km\,s$^{\rm -1}$. The calibrated visibilities were imaged to match the spatial and spectral resolution of the Band~5 data.

\subsubsection{Contribution of line blending} \label{subsec:obs_synthetic_spec}

Since the mm spectrum of Arp~220 contains numerous spectral lines \citep[see e.g.,][]{mar11}, we carefully investigate the potential contamination 
by other species. Both the 183 and 325~GHz water lines are affected by blending with other molecular transitions. To correctly estimate the water 
line intensities, taking into account this contribution from other species, we used the synthetic spectrum model fit to Band~6 and 7 observations 
(Mart\'{i}n et al. in prep.). The model fit was performed assuming LTE conditions and includes the contribution from more than 30 species, 
isotopologues and vibrationally excited states.\\
\indent
At 183~GHz we use the synthetic spectrum and extrapolate the model fit to Band~5. Fig.\,\ref{fig:spectra} shows that the extrapolation to Band~5 
appears to properly fit the emission outside the water line. The main contribution to the observed profile around this water line is the emission 
of both C$_{\rm 2}$H$_{\rm 5}$CN and HC$_{\rm 3}$N~v$_{\rm 7}$\,=\,2 at 183.1~GHz and the c-C$_{\rm 3}$H$_{\rm 2}$ at 183.6~GHz. The main 
contaminants to the 325~GHz H$_{\rm 2}$O line are CH$_{\rm 2}$NH emission at 325.3~GHz and to a lesser extent CH$_{\rm 3}$CCH emission at 324.6~GHz.\\
\indent
All flux values for the interferometrically observed water lines stated from this point on refer to the spectra corrected for their contaminants, 
and should thus only contain contributions by the corresponding water lines.

\subsection{IRAM 30~m} \label{subsubsec:obs_iram}

IRAM 30m observations of the H$_{\rm 2}$O\,3$_{\rm 13}-2_{\rm 20}$ and H$_{\rm 2}$O\,5$_{\rm 15}-4_{\rm 22}$O lines 
($\nu$$_{\rm rest}$\,=\,183.310~GHz, and 325.153~GHz respectively) towards the center of Arp~220 were performed in December 2012 (project: 189-12, 
PI: J. Cernicharo) and April/May 2014 (project: 186-13; Krips et al., in prep.). The beam sizes of the IRAM 30~m telescope at these frequencies are 
14\arcsec\ and 8\arcsec. We used EMIR in conjunction with the FTS backend at a spectral resolution of 0.2~MHz over a bandwidth of 4~GHz 
($\sim$4000~km\,s$^{\rm -1}$ and $\sim$6750~km\,s$^{\rm -1}$, respectively). Regular pointing and focus measurements on bright nearby quasars and 
a line source to verify the absolute flux calibration were conducted. Typical system temperatures were T$_{\rm sys}$$\simeq$300-1400~K at 2~mm and 
T$_{\rm sys}$$\simeq$300-1200~K at 0.8~mm\footnote{The high T$_{\rm sys}$ values are due to the fact that the edges of the observed frequency bands 
are close to the atmospheric water absorption line around 183~GHz and 325~GHz.}. We converted T$_{\rm A}^*$ scales using 
T$_{\rm mb}$\,=\,T$_{\rm A}^*\times$F$_{\rm eff}$/B$_{\rm eff}$ with the following efficiencies taken from the IRAM 30m webpage: 
F$_{\rm eff}$\,=\,0.93 and B$_{\rm eff}$\,=\,0.74 at 180~GHz and F$_{\rm eff}$\,=\,0.81 and B$_{\rm eff}$\,=\,0.35 at 320~GHz sky frequency.

\subsection{22~GHz VLA data} \label{subsec:obs_22ghz}

The VLA H$_{\rm 2}$O\,6$_{\rm 16}-5_{\rm 23}$ line at 22.235~GHz rest frequency is blended with the NH$_{\rm 3}$\,(3,1) absorption line, 
with a major effect towards Arp~220 West \citep[see Fig.3 in][]{zsch16}. In order to retrieve the intrinsic profiles of the 22~GHz water 
line, we simultaneously fit the NH$_{\rm 3}$ lines within the VLA spectrum closest to the water line: NH$_{\rm 3}$\,(3,1) (rest frequency: 
22.235~GHz) and NH$_{\rm 3}$\,(4,2) (rest frequency: 21.703~GHz). We assume a simple Gaussian line profile, with the same centroid velocity 
and linewidth for both lines, and a relative intensity scaling between NH$_{\rm 3}$\,(4,2) and NH$_{\rm 3}$\,(3,1). We find 
v$_{\rm 0}$\,= -94.5\,$\pm$\,5.5~km\,s$^{\rm -1}$, FWHM\,=\,287\,$\pm$\,13~km\,s$^{\rm -1}$, and a line ratio 
NH$_{\rm 3}$\,(4,2)-to-NH$_{\rm 3}$\,(3,1) of 1.14\,$\pm$\,0.08. A comparison of the spectra before and after the fit is shown in 
Fig.\,\ref{fig:22ghz_comp}. For Arp~220 East, the NH$_{\rm 3}$ absorption is weak relative to the H$_{\rm 2}$O line. Thus we keep the original 
water spectrum without trying to correct for NH$_{\rm 3}$ absorption. To be able to compare these data to the water lines at 183 and 325~GHz, 
the spatial and spectral resolution were degraded to the coarser resolution of the Band~5 data ($\sim$20~km\,s$^{\rm -1}$, 
0.79\arcsec\,$\times$\,0.71\arcsec).

%__________________________________________________________________

\section{Results} \label{sec:results}

\begin{table*}[ht]%[!h]
\begin{minipage}[!h]{\textwidth}
\centering
\renewcommand{\footnoterule}{}
\caption{\small
 Water line properties.}
\label{tab:water_properties}
\tabcolsep0.1cm
\begin{tabular}{lcccccccc}
\hline
\noalign{\smallskip}
\hline
\noalign{\smallskip}
%---------------------------------------------------------------------------
Line & $\nu$$_{\rm rest}$ & E$_{\rm upper}$ & Telescope & Beam & \multicolumn{2}{c}{$\int$$S$\,d$v$}\footnote{When two values are given, 
the left column denotes the contamination-corrected integrated intensities measured in Arp~220 East, the right \newline \hspace*{3mm} column is 
for Arp~220 West. A single value given represents the uncorrected integrated intensities for a region that encompasses both nuclei.}\footnote{The 
integrated intensities given for the interferometric observations are the contamination-corrected values.} & Date of & Reference
\footnote{References: (1) \citet{zsch16}, (2) \citet{cer06}, (3) this work, (4) \citet{gal16}.} \\
\noalign{\smallskip}
     & [GHz]              & [K]         &   &      & \multicolumn{2}{c}{[Jy\,km\,s$^{\rm -1}$]} & observation &           \\
%---------------------------------------------------------------------------
\noalign{\smallskip}
\hline
\noalign{\smallskip}
H$_{\rm 2}$O\,6$_{\rm 16}-5_{\rm 23}$ & 22.23508  & 609 & VLA       & 0.79\arcsec\,$\times$\,0.71\arcsec & 0.34 & 0.17              & March 2011-June 2012 & 1 \\
H$_{\rm 2}$O\,3$_{\rm 13}-2_{\rm 20}$ & 183.31009 & 205 & IRAM 30~m & 14\arcsec                          & \multicolumn{2}{c}{55.7\footnote{Derived from the luminosity given in their Table\,1.}}  & January 2005         & 2 \\
                                      &           &     & IRAM 30~m & 14\arcsec                          & \multicolumn{2}{c}{56.9} & April, May 2014 & 3 \\
                                      &           &     & APEX      & 35\arcsec                          & \multicolumn{2}{c}{50\,$\pm$\,6} & July, September 2015 & 4 \\
                                      &           &     & ALMA      & 35\arcsec                          & \multicolumn{2}{c}{56.0} & July 2016            & 3 \\
                                      &           &     & ALMA      & 0.79\arcsec\,$\times$\,0.71\arcsec & 13.7 & 47.8              & July 2016            & 3 \\
H$_{\rm 2}$O 5$_{\rm 15}-4_{\rm 22}$  & 325.15292 & 470 & IRAM 30~m & 8\arcsec                           & \multicolumn{2}{c}{39.8} & December 2012   & 3 \\
                                      &           &     & IRAM 30~m & 8\arcsec                           & \multicolumn{2}{c}{42.3} & April, May 2014 & 3 \\
                                      &           &     & ALMA      & 8\arcsec                           & \multicolumn{2}{c}{45.4} & June, July 2014      & 3 \\
                                      &           &     & ALMA      & 0.79\arcsec\,$\times$\,0.71\arcsec & 9.1 & 25.3               & June, July 2014      & 3 \\
\noalign{\smallskip}
\hline
%--------------------------------------------------------------------------
\end{tabular}
\end{minipage}
\end{table*}

\begin{figure*}[ht]
  \begin{minipage}[hbt]{0.24725\textwidth}
  \centering
    \includegraphics[width=\textwidth]{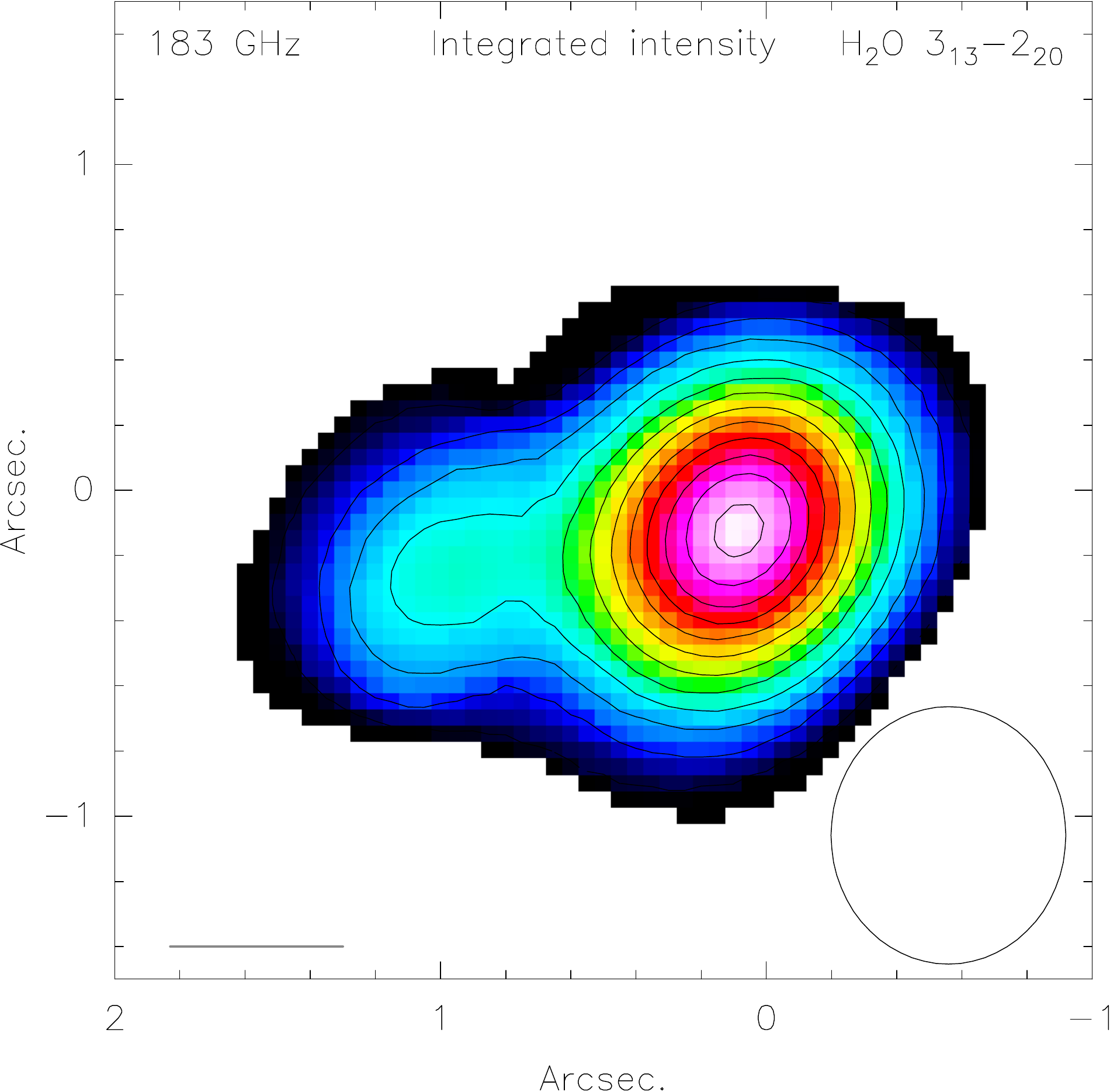}
  \end{minipage}
  \begin{minipage}[hbt]{0.24725\textwidth}
  \centering
    \includegraphics[width=\textwidth]{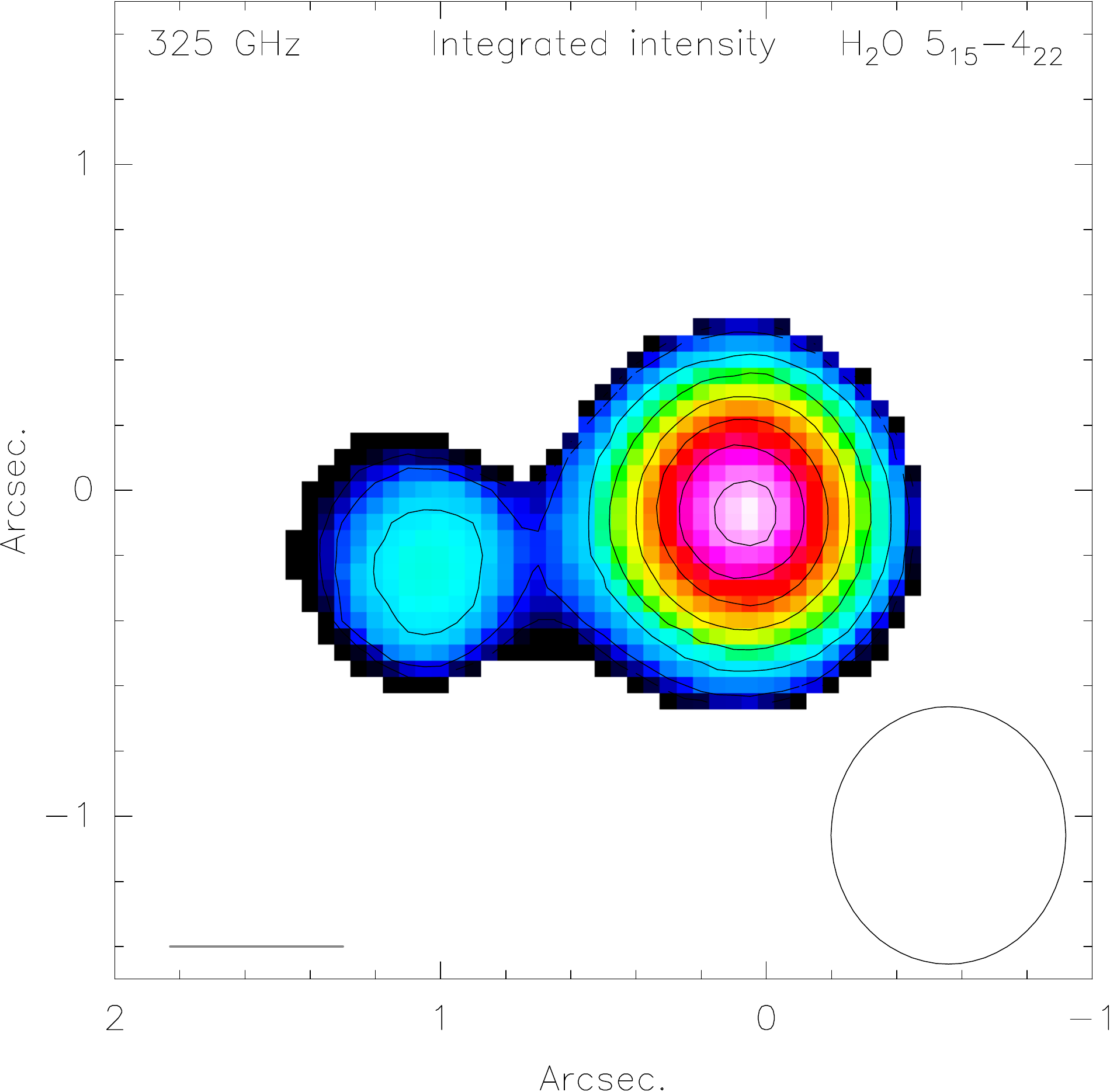}
  \end{minipage}
  \begin{minipage}[hbt]{0.24725\textwidth}
  \centering
    \includegraphics[width=\textwidth]{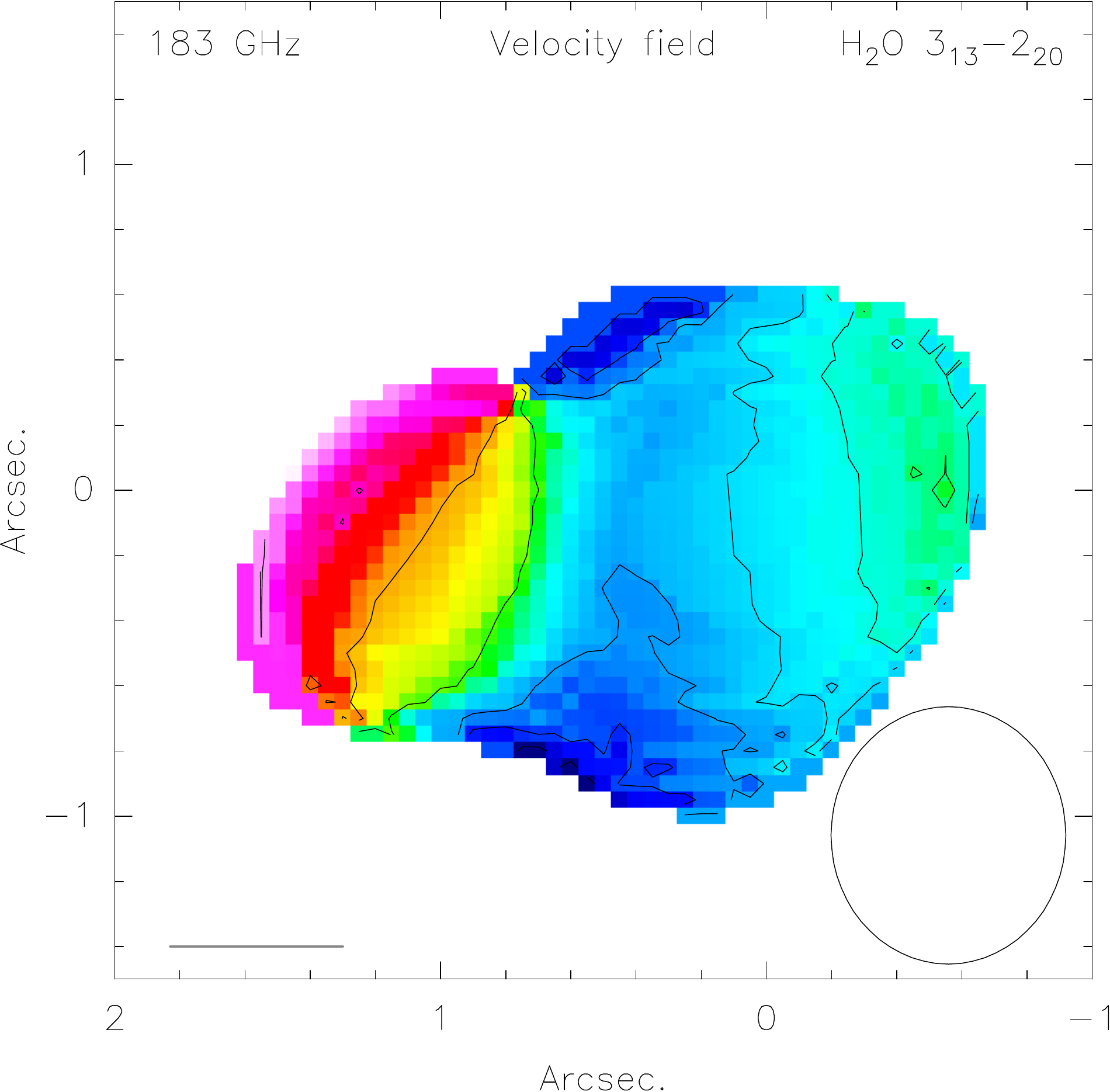}
  \end{minipage}
  \begin{minipage}[hbt]{0.24725\textwidth}
  \centering
    \includegraphics[width=\textwidth]{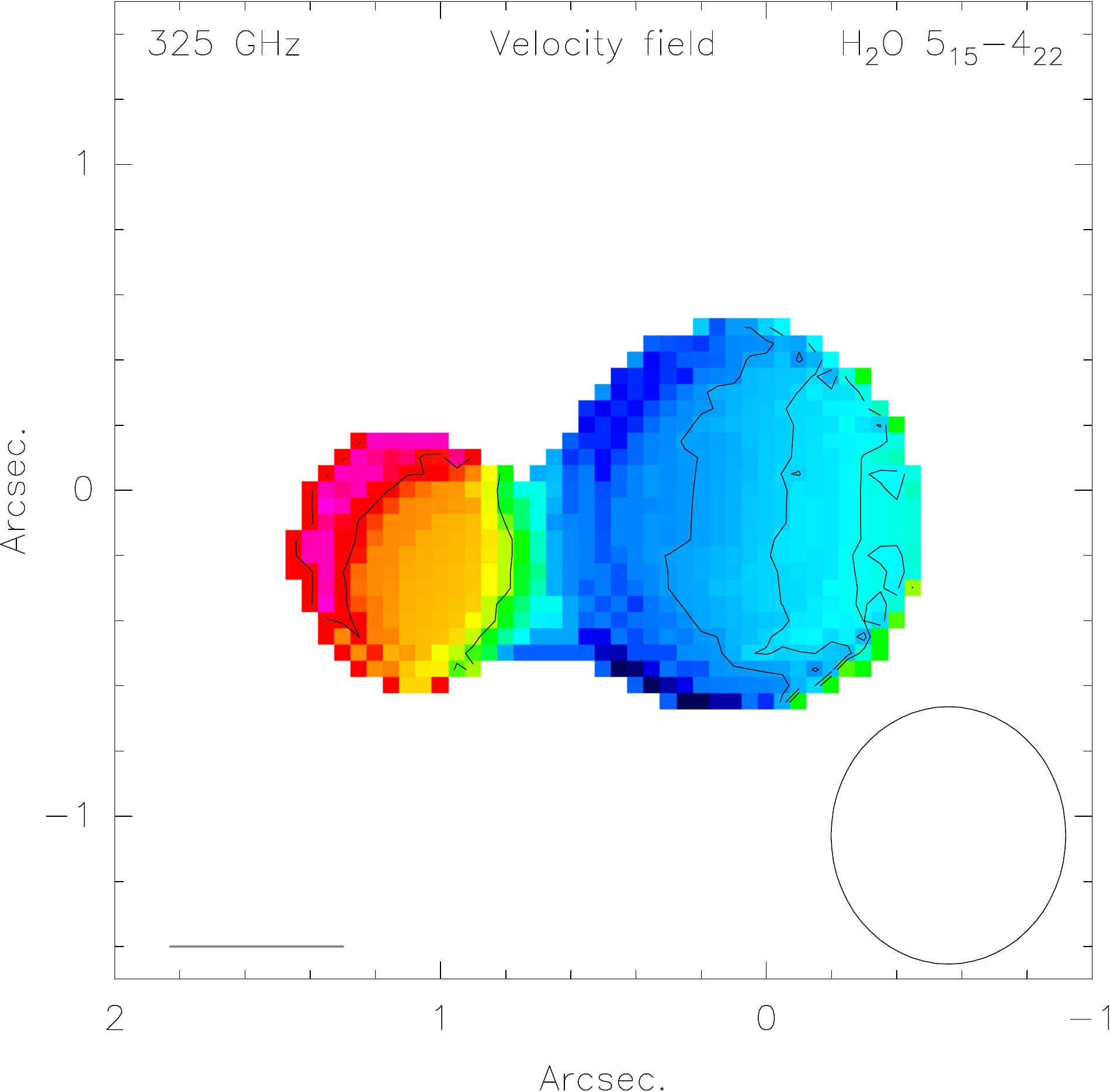}
  \end{minipage}
  \caption{\footnotesize Maps of the integrated intensity distributions (\textit{left}) and the velocity fields (\textit{right}) of the 183 
   and 325~GHz water lines. \textit{Integrated intensity maps:} The phase center is located at Arp~220 West ($\alpha$=15:34:57.220, 
   $\delta$=+23:30:11.60). The contours start at 5$\sigma$ and are spaced in steps of 10$\sigma$ 
   (1$\sigma$(183~GHz)\,=\,3.9~mJy\,km\,s$^{\rm -1}$, 1$\sigma$(325~GHz)\,=\,5.4~mJy\,km\,s$^{\rm -1}$). \textit{Velocity fields:} The colour 
   scale is identical for both lines (5200~km\,s$^{\rm -1}$-5600~km\,s$^{\rm -1}$), the velocity contours start at 5280~km\,s$^{\rm -1}$ and 
   are spaced in steps of 30~km\,s$^{\rm -1}$. North is up, east to the left. The beam (0.79\arcsec\,$\times$\,0.71\arcsec) is indicated in the 
   bottom right corner of each image. The bar in the lower left corner represents a spatial scale of 200~pc.}
  \label{fig:maps}
\end{figure*}

\subsection{Water} \label{subsec:results_water}

\textit{183~GHz:} The emission from the H$_{\rm 2}$O\,3$_{\rm 13}-2_{\rm 20}$ line is detected in both nuclei (Figs.\,\ref{fig:spectra}, 
\ref{fig:maps}). The brightest flux is found in Arp~220 West (integrated intensity: $\sim$47.8~Jy\,km\,s$^{\rm -1}$). Arp~220 East is about three 
times less luminous ($\sim$13.7~Jy\,km\,s$^{\rm -1}$). The velocity field shows a much steeper gradient in Arp~220 East than in Arp~220 West 
(Fig.\,\ref{fig:maps}).\\
\noindent
\textit{325~GHz:} For the first time we detect the 325~GHz H$_{\rm 2}$O\,5$_{\rm 15}-4_{\rm 22}$ line in an extragalactic environment. Both, 
single-dish (IRAM 30~m) and interferometric observations (ALMA), led to detections of this water transition 
(Figs.\,\ref{fig:spectra},\ref{fig:maps},\ref{fig:variability}, see also Table\,\ref{tab:water_properties}). Similar to the 183~GHz line, 
325~GHz emission is found in both nuclei, with Arp~220 West being the brightest. Integrated fluxes amount to $\sim$25.3~Jy\,km\,s$^{\rm -1}$ 
and 9.1~Jy\,km\,s$^{\rm -1}$ in Arp~220 West and East, respectively. The velocity field (Fig.\,\ref{fig:maps}) shows a similar behaviour between 
Arp~220 East and Arp~220 West as at 183~GHz.\\
\noindent
\textit{22~GHz:} The correction of the H$_{\rm 2}$O\,6$_{\rm 16}-5_{\rm 23}$ emission line for the NH$_{\rm 3}$\,(3,1) absorption leaves us with a 
picture opposite to what we find for the 183 and 325~GHz water lines: The 22~GHz emission is brightest in Arp~220 East (Fig.\,\ref{fig:spectra}, 
integrated intensity: $\sim$0.34~Jy\,km\,s$^{\rm -1}$). The emission in Arp~220 West, however, is quite faint ($\sim$0.17~Jy\,km\,s$^{\rm -1}$), 
which is quite surprising. One caveat the NH$_{\rm 3}$\,(3,1) leaves us with, is that the two emission peaks in Arp~220 East could actually be not 
two separate peaks but one broad line with the NH$_{\rm 3}$ absorption on top of the water line. If that is the case, then the 22~GHz water line in 
Arp~220 East is even brighter than what we take into account here. This could also mean that the majority of the 22~GHz line in Arp~220 West is 
absorbed so that we are missing this velocity component, which would also explain the apparent misalignment of the velocities between the eastern and 
western nuclei visible in Fig.\,\ref{fig:spectra}. A contribution of the $^{\rm 13}$CH$_{\rm 3}$OH line as mentioned in \citet{zsch16} is very 
unlikely; this line has not been detected in any of the broad frequency spectral scans conducted in Arp~220 so far \citep[e.g.,][]{mar11,ala15}.\\

\subsection{Other identified lines in Band~5} \label{subsec:results_other_lines}

Other detected lines included in the Band~5 tuning are HNC\,2$-$1, CH$_{\rm 3}$OH\,4$-$3 and CS\,4$-$3 (see Fig.\,\ref{fig:spec_b5}). These lines 
have been previously observed with SEPIA at APEX by \citet{gal16}. Furthermore, fainter lines identified within the respective frequency 
ranges are vibrationally excited HC$_{\rm 3}$N (several lines close to H$_{\rm 2}$O\,3$_{\rm 13}-2_{\rm 20}$) and CH$_{\rm 3}$CN\,10$-$9 (see also 
Table\,\ref{tab:properties_band_5_lines}), as well as potentially H$^{\rm 13}$CCCN, C$^{\rm 34}$S\,4$-$3 and $^{\rm 13}$CH$_{\rm 3}$CN. They have 
been identified using the synthetic spectrum extrapolated from Band~6 and 7 (Mart\'{i}n et al., in prep.) to Band~5 (see Sect.\,\ref{sec:obs}). Most 
of these lines are fainter in Arp~220 East than in Arp~220 West. They all have been previously detected at 1 and/or 3~mm by e.g., \citet{mar11} and 
\citet{ala15}.

%______________________________________________________________

\section{Discussion} \label{sec:discussion}

In 2006, \citeauthor{cer06} observed the H$_{\rm 2}$O\,3$_{\rm 13}-2_{\rm 20}$ line at 183~GHz rest frequency for the first time in Arp~220. 
Its isotropic luminosity placed it firmly in the range of megamasers. This detection was later confirmed by \citet{gal16}. The absence of 
time variations in the line in-between the two observing periods and the similar line profile to H$_{\rm 2}$$^{\rm 18}$O \citep{mar11} led to 
speculations about the origin of the water emission: both analyses favor thermal processes and/or a large number of star-forming cores as the 
cause of the observed water signature, rather than an AGN maser origin. However, at the time \citet{cer06} performed the modelling only information 
about the spatially unresolved 183~GHz line were available. The authors concluded that only a combination of the three major water lines at 22, 
183 and 325~GHz will provide a clear picture of the excitation conditions in the molecular gas emitting water emission in Arp~220. Their 
excitation analysis was further limited by the lack of spatial resolution. Hence, this work, where we make use of subarcsecond-scale observations 
of all three lines, does represent a major step forward to find the true powering source of the bright water emission in Arp~220.

\subsection{Time variability}

\begin{figure}[t]
  \centering
    \includegraphics[width=0.4\textwidth]{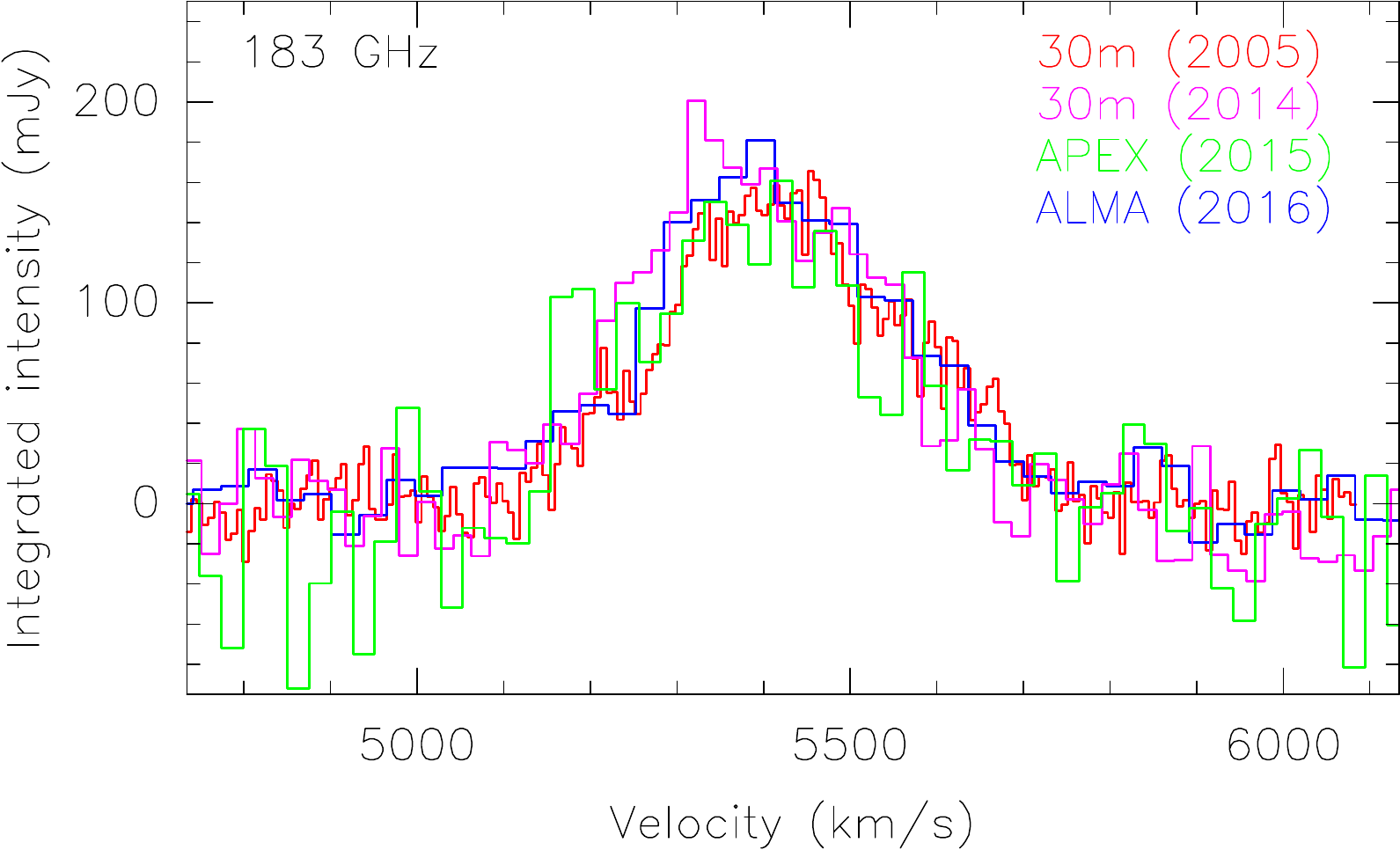}
    \includegraphics[width=0.4\textwidth]{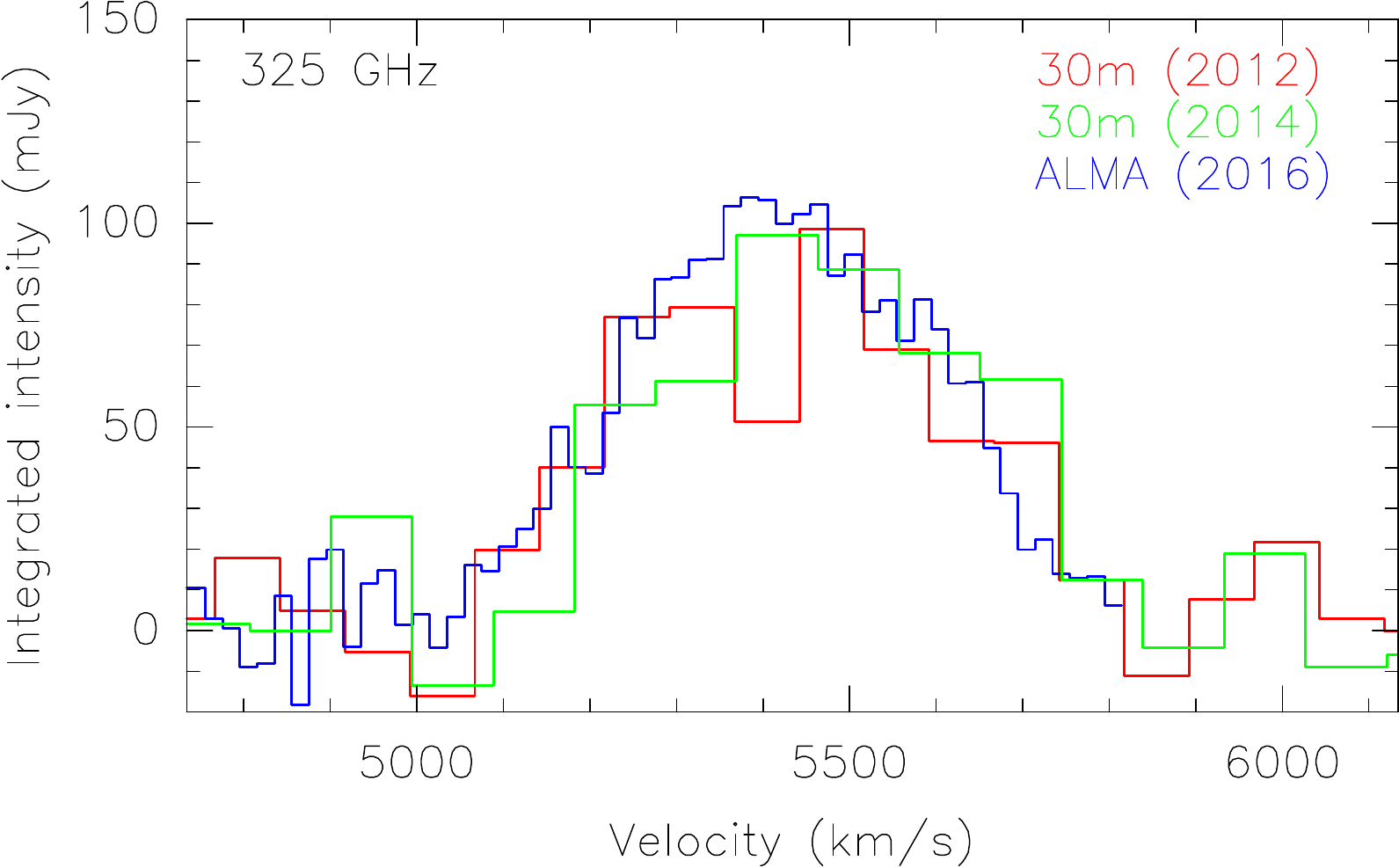}
  \caption{\footnotesize Comparison of the intensities and line shapes between previous single-dish observations (IRAM 30~m, APEX) and spatially 
   smoothed ALMA observations at 183 and 325~GHz. The spectra are in good agreement within the errors imposed by the differences in the 
   sensitivity of the used instruments.}
  \label{fig:variability}
\end{figure}

We compare the line contamination uncorrected ALMA Band~5 data with previous single-dish spectra from the IRAM 30~m and APEX (see 
Table\,\ref{tab:water_properties}). A degradation of the convolving beam allows us to compare to what has been found by \citet{cer06} and 
\citet[][Fig.\,\ref{fig:variability}]{gal16}. The flux recovered from the degraded ALMA data (35\arcsec\ aperture) is 
$\sim$56.0~Jy\,km\,s$^{\rm -1}$. This is 10\% larger than the flux found by \citet{gal16}, which matches their measurement uncertainties. Line 
shapes and widths (FWHM) are in agreement as well: 318~km\,s$^{\rm -1}$ vs. 310~km\,s$^{\rm -1}$ from \citet{cer06}, and 
332\,$\pm$\,31~km\,s$^{\rm -1}$ from \citet{gal16}.\\
\indent
We also degraded the ALMA 325~GHz data to the same beam size as the IRAM 30~m observations (8\arcsec, Fig.\,\ref{fig:variability}). The integrated 
intensity determined from the ALMA data ($\sim$45.4~Jy\,km\,s$^{\rm -1}$) is slightly higher than what we find for the IRAM 30~m observations 
($\sim$39.8~Jy\,km\,s$^{\rm -1}$ in December 2012, and $\sim$42.3~Jy\,km\,s$^{\rm -1}$ in April/May 2014). However, the signal-to-noise ratio in 
the IRAM data was quite low compared to the higher sensitivity ALMA data set, which also necessitated a quite heavy spectral binning which made 
the determination of the FWHM line width difficult. The extracted FWHM are $\sim$440~km\,s$^{\rm -1}$ for ALMA and $\sim$445~km\,s$^{\rm -1}$ for 
the IRAM 30~m spectra (Fig.\,\ref{fig:variability}).\\
\indent
The shortest baselines in the ALMA observations correspond to spatial frequencies of $\sim$11\arcsec\ (Band~5) and $\sim$14\arcsec\ (Band~7). Thus 
the interferometer should recover all structures $\leq$5\arcsec\, that are larger than the size of the two nuclei 
\citep[e.g.,][]{dow07,aal09,sak09,mar11}. The consistency between the contamination-uncorrected ALMA Band~5 and single-dish spectra (intensity, 
line shape, see Fig.\,\ref{fig:variability}) suggests that the water line did not vary in time in between observations. The lack of 
variability in the lines gives us further confidence that the line ratios used in the LVG modeling below, are robust despite being observed at 
different times. Typical timescales on which those variations are expected for extragalactic water megamaser emission can range from minutes to 
days to weeks to months to years \citep[e.g.,][and references therein]{gre97,ral98,bra03,lo05}.

\subsection{Line morphology} \label{subsec:line_morphology}

\begin{figure}[t]
  \centering
    \includegraphics[width=0.4\textwidth]{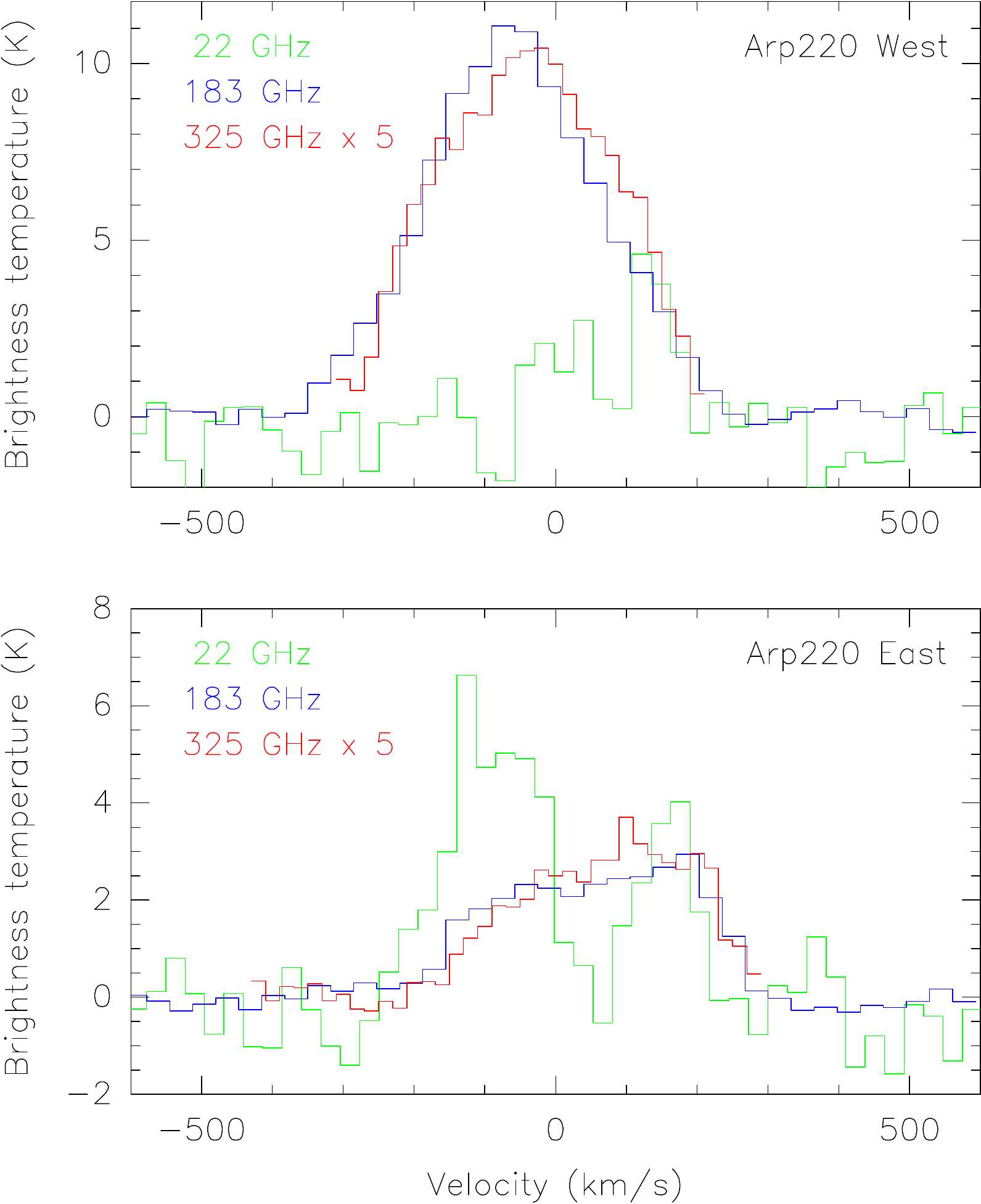}
  \caption{\footnotesize Comparison of the line shapes and peak temperatures of the contamination-corrected water spectra at 22 (in green), 183 
   (blue) and 325~GHz (red) in Arp~220 West (\textit{top}) and East (\textit{bottom}).}
  \label{fig:comp_line_shape}
\end{figure}

\subsubsection{Ratios} \label{subsubsec:ratios}

A comparison of the integrated intensity values given in Sect.\,\ref{subsec:results_water} shows that the integrated intensity 183-to-325~GHz ratios 
for the two nuclei, I$_{\rm West}$\,(183GHz)/I$_{\rm West}$\,(325GHz) and I$_{\rm East}$\,(183GHz)/I$_{\rm East}$\,(325GHz), are 1.9 and 1.5, 
respectively. The intensity ratios between the nuclei at each of the two frequencies differ as well: 
I$_{\rm 183}$\,(West)/I$_{\rm 183}$\,(East)\,$\sim$3.5 and I$_{\rm 325}$\,(West)/I$_{\rm 325}$\,(East)\,$\sim$2.8.\\
\indent
The observed peak brightness temperatures at 22, 183 and 325~GHz are comparable: 6.6~K at 22~GHz (in Arp~220 East, see also 
Table\,\ref{tab:peak_brightness+ratios}), 11.1~K at 183~GHz (in Arp~220 West), 2.1~K at 325~GHz (in Arp~220 West). These temperatures are much lower 
than what is typically expected for maser emission \citep[$\sim$10$^{\rm 9-10}$~K or more;][and references therein]{sly03,cer06b,ces08,gra12}. Thus, 
despite the high isotropic luminosity, and together with the lack of variability at 183 and 325~GHz it seems unlikely that the water emission in 
Arp~220 could have a maser origin if coming from one large cloud (e.g., of the size of the beam) surrounding the nuclei. A thermal origin was 
suggested as an alternative by \citet{mar11} and \citet{gal16}. However, thermal emission from the three lines seems unlikely as the 22~GHz line 
arises from an energy level E$_{\rm upper}$ at 609~K (see Table\,\ref{tab:water_properties}) which will require very specific physical conditions to 
have pure thermal emission. Moreover, the observed line intensity ratios are not compatible with pure thermal emission. Hence, it is very likely 
possible that the emission region is much smaller than the beam size we reach in this work, e.g., a large number of smaller clumps emitting in the 
water lines. Consequently, the corresponding brightness temperatures would rise considerably above the values measured here. High angular 
resolution observations with the most extended configurations of ALMA are needed to determine the true projected surface area of the H$_{\rm 2}$O 
emission.

\subsubsection{Line shapes} \label{subsubsec:line_shape}

Compared to extragalactic maser emission, e.g., in Circinus \citep[e.g.,][]{hag13,hag16,pes16}, the line shapes of the 183~GHz and 325~GHz water lines 
in Arp~220 are rather unspectacular. The complex velocity structure with multiple narrow features commonly associated with megamaser emission is not 
apparent in Arp~220. The contamination-free spectra show broad emission lines (FWHM $\sim$318~km\,s$^{\rm -1}$) that are uniform in their structure 
(Fig.\,\ref{fig:comp_line_shape}). In Arp~220 East, the emission lines have a boxy, double-peaked, non-Gaussian appearance. In Arp~220 West the 
spectra are single-peaked symmetric and Gaussian-like. This is also corroborated in the velocity fields at both frequencies (Fig.\,\ref{fig:maps}). 
Extragalactic sources exhibiting maser lines of complex velocity structure are typically those where the maser is associated with the accretion disk 
or torus close to an AGN. These so-called \textit{disk-masers} are mostly found in Seyfert 2 and LINER galaxies 
\citep[low ionization emission region, e.g.,][]{kar99,kuo11,kuo15}. In other galaxies megamaser emission is caused by the interaction between the 
AGN jet and molecular clouds in the surrounding interstellar medium \citep[\textit{jet-masers}, e.g., Mrk~348;][]{pec03,cas16}. The maser lines in 
these sources are typically as featureless as what we find in Arp~220. However, a comparison of the velocity distributions in the water emission 
in Arp~220 to e.g., the CO\,1$-$0 velocity field \citep[e.g.,][]{sco16} shows a similar rotation pattern. Thus it seems unlikely that the maser 
emission is due to an interaction between an AGN jet and molecular clouds in its vicinity. The signature of this process would introduce a 
disturbance in the velocity field that would be apparently different from the rotation pattern observed at the spatial and spectral resolutions in 
the data. A complementary scenario previously suggested to explain the maser emission in Arp~220 is that it originates from a large number of 
star-forming cores, comparable to what has been found in \object{Sgr~B2} \citep{cer06,gal16}. Indeed, high-resolution VLBI observations have 
discovered almost fifty point sources in the central 0.5~kpc that have been identified as radio supernovae (RSNe) and supernova remnants 
\citep[SNRs,][]{smi98,lon06,par07,bat11}, pointing towards very active star formation. In fact, with a star formation rate of few 
100~M$_{\sun}$\,yr$^{\rm -1}$ \citep[e.g.,][]{ana00,thr08,var16}, it is certainly possible that the brightness of water lines observed in Arp~220 is  
entirely due to the emission from a large number of star-forming regions.

\subsection{Excitation}

\begin{table}[t]%[!h]
\centering
\renewcommand{\footnoterule}{}
\caption{\small
 Peak brightness temperatures and their ratios.}
\label{tab:peak_brightness+ratios}
\tabcolsep0.1cm
\begin{tabular}{lccc}
\hline
\noalign{\smallskip}
\hline
\noalign{\smallskip}
%---------------------------------------------------------------------------
 & East & West & West/East \\
%---------------------------------------------------------------------------
\noalign{\smallskip}
\hline
\noalign{\smallskip}
 22~GHz  & 6.6~K  & 4.6~K  & 0.7 \\
 183~GHz & 2.9~K  & 11.1~K & 3.8 \\
 325~GHz & 0.74~K & 2.1~K  & 2.8 \\
\noalign{\smallskip}
\hdashline
\noalign{\smallskip}
 183/325 & 3.9  & 5.3  &  \\
 183/22  & 0.44 & 2.4  &  \\
 325/22  & 0.11 & 0.46 &  \\
\noalign{\smallskip}
\hline
%--------------------------------------------------------------------------
\end{tabular}
\end{table}

\begin{figure}[t]
  \centering
    \includegraphics[width=0.48\textwidth]{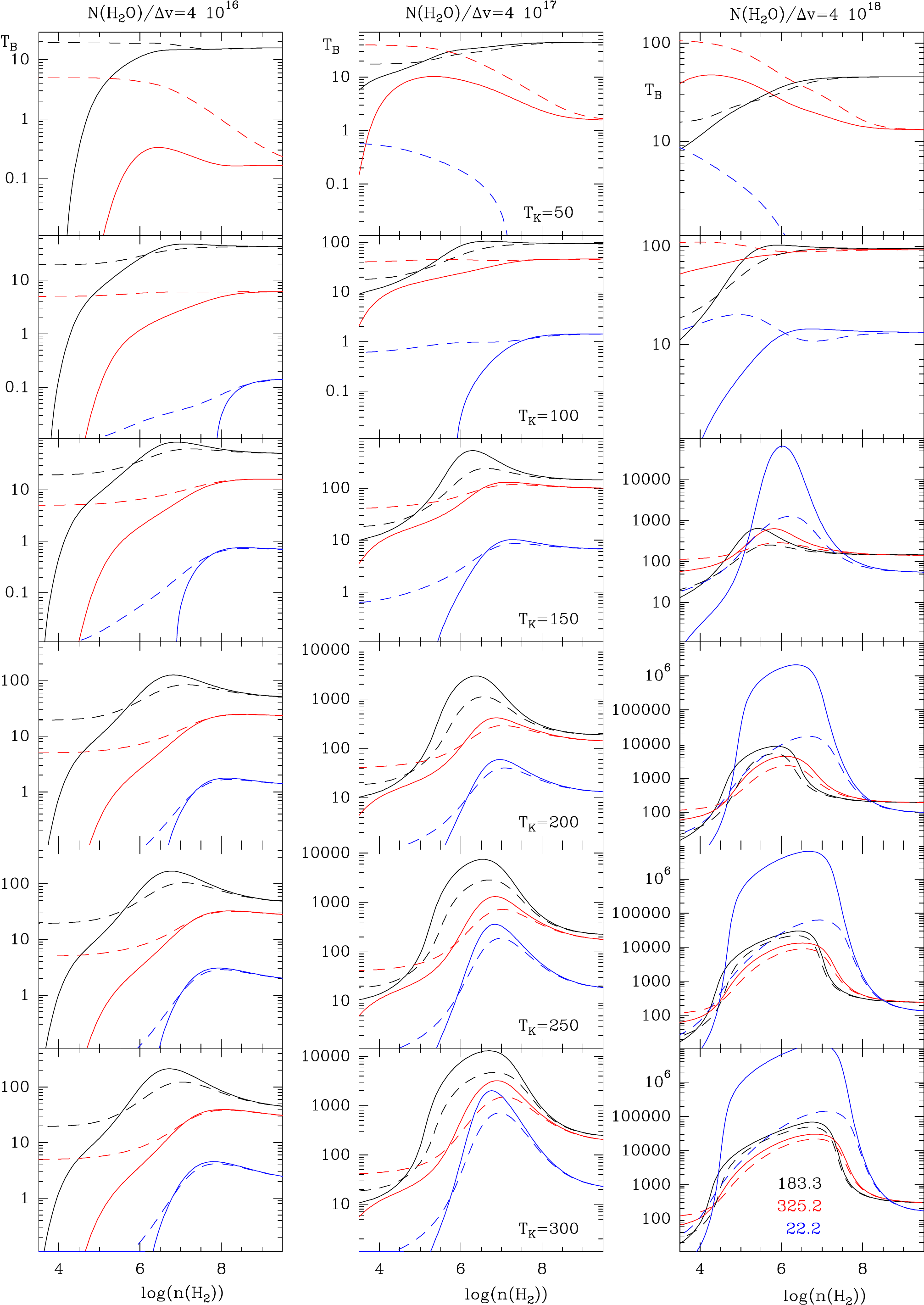}
  \caption{\footnotesize LVG models covering a wide range of physical conditions for three different column densities. For six different 
   kinetic temperatures (T$_{\rm K}$), the gas temperature variation over a range of H$_{\rm 2}$ densities is shown. The blue, black and red colours 
   represent the results for the 22, 183 and 325~GHz lines, respectively. Solid lines represent the expected intensities without continuum emission 
   while dashed lines show the results when infrared pumping from dust is taken into account.}
  \label{fig:modelling}
\end{figure}

In the following discussion, we assume that the observed 22~GHz emission is real and that its observed intensity has an uncertainty of 50\% due to 
the strong contamination by NH$_{\rm 3}$. The three lines observed in this paper arise from upper energy levels at 205~K (183~GHz), 470~K (325~GHz), 
and 609~K (22~GHz; the given levels correspond to the ortho species which is 34~K above the para one) and have Einstein A coefficients of 
3.62\,$\times$\,10$^{\rm -6}$~s$^{\rm -1}$, 1.15\,$\times$\,10$^{\rm -5}$~s$^{\rm -1}$ and 1.98\,$\times$\,10$^{\rm -9}$~s$^{\rm -1}$ respectively. 
Hence, excitation conditions could be very different for the three lines and the observed emission could trace different regions of the clouds. A 
comparison of the expected brightness temperatures of the three lines at 22, 183 and 325~GHz in Arp~220 was previously made by \citet{cer06}. We 
here repeat these calculations using new collisional rates of \citet{dan11} including dust emission excitation effects on the population of the 
rotational levels of H$_{\rm 2}$O. In all cases we assumed that the main collider is ortho H$_{\rm 2}$. Dust emission will affect the range of 
volume and column densities under which these lines will have a significant emission in Arp~220. For the far-IR lines of water vapor observed in 
this source \citet{gon04,gon12} concluded that all the lines were formed around the optically thick continuum sources in the two nuclei of 
Arp~220. The dust opacity is so large that these lines appear in absorption as they are sub-thermally excited. However, the 22, 183 and 325~GHz 
lines observed here appear in emission and collisional excitation does play an important role in the emerging intensity. Moreover, unlike the 
far-IR lines that can not give information about regions of high dust opacity, the three sub-millimeter, millimeter and radio lines considered here 
could transport some information from these regions.\\
\indent
We have modelled the water emission at 22, 183 and 325~GHz in Arp~220 using the MADEX code \citep{cer12}. The model includes a central source of 
dust emission to assess the effect of the dust on the expected intensity of the lines. Fig.\,\ref{fig:modelling} shows the predicted intensities for 
a large range of H$_{\rm 2}$ densities, gas temperatures (50 to 300~K) and column densities N(H$_{\rm 2}$O)/$\Delta$v 
(4$\times$10$^{\rm 16}$ to 4$\times$10$^{\rm 18}$~cm$^{\rm -3}$\,(km\,s$^{\rm -1}$)$^{\rm -1}$). As for other water maser models, such as 
e.g., \citet{neu91}, we assume an ortho-to-para ratio of 3:1. Solid lines represent the expected intensities without continuum emission while dashed 
lines show the results when infrared pumping from dust is taken into account. The central source has a size of 5~pc and the layer of water vapor was 
placed at a distance of 1.5~pc from the outer radius of the continuum source. As expected, the effect of infrared pumping is a reduction the maser 
effect (in particular for the 22~GHz line) and a shift of the intensity maximum towards higher densities. For low gas densities and temperatures, 
the presence of infrared pumping increases significantly the emission of the three water lines. The predicted brightness temperatures, in absence of 
dust, agree very well with those of \citet{cer06}. As the density and gas temperature increase, masing effects do appear in a similar way than what 
has been found in their work. Placing the layer of water vapor at larger distances from the continuum dust emission source reduces the infrared 
pumping considerably.\\
\indent
To pinpoint the range of parameters describing the physical conditions in the emission regions of the water lines Arp~220, we start from the most 
simple assumption: the water emission at 22, 183 and 325~GHz arises from one single large cloud. If a continuum source is not considered, the 
observed intensities of the 183~GHz and 325~GHz lines alone could be reproduced simultaneously within a parameter space of 
n(H$_{\rm 2}$)\,=\,10$^{\rm 4}$-10$^{\rm 5}$~cm$^{\rm -3}$, T$_{\rm K}$\,=\,100~K, and 
N(H$_{\rm 2}$O)/$\Delta$v\,=\,4\,$\times$\,10$^{\rm 17}$~cm$^{\rm -3}$\,(km\,s$^{\rm -1}$)$^{\rm -1}$. If dust emission is taken into account, no 
solution can be found for the 183 and 325~GHz lines. On top of that, both these models fail to reproduce the 22~GHz line. It is thus not possible to 
converge on a solution for this most simple scenario.\\
\indent
Therefore, our next approach is to model the water emission as originating from several smaller clouds. To estimate the impact of the number and size 
of the clouds, we introduce a dilution factor d. The dilution factor is a measure of the area occupied by a structure of radius r inside the region 
covered by the beam: (r/r$_{\rm beam}$)$^{\rm 2}$ (for one source). For N sources, the dilution factor will be 
$\sum$r$_{\rm i}^{\rm 2}$/r$_{\rm beam}$$^{\rm 2}$. A dilution factor of 1 means that the solid angle of the source is equal to the solid angle of the 
beam. If d is 0.1, 10\% of the beam area is filled by the source(s). A d of 0.001 means that the emission is coming from a small fraction of the beam. 
The dilution factor also has an influence on the true brightness temperature -- T$_{\rm B}$/d.\\
\indent
Assuming that the water emission originates from a number of smaller clouds does indeed result in a common solution for all three lines -- for large 
dilution factors d at high column densities (N(H$_{\rm 2}$O)/$\Delta$v\,=\,4\,$\times$\,10$^{\rm 18}$~cm$^{\rm -3}$\,(km\,s$^{\rm -1}$)$^{\rm -1}$) 
and high temperatures (T$_{\rm K}$\,=\,200-300~K). If no continuum source is present, the dilution factors we have to apply are 0.005 for 
T$_{\rm K}$\,=\,200~K and 0.0025 for T$_{\rm K}$\,=\,300~K. The density for which this solution is found is rather low, around 
8\,$\times$\,10$^{\rm 4}$~cm$^{\rm -3}$. If the continuum source is present, the dilution factor has to be 0.001 for T$_{\rm K}$=\,200~K and 0.0007 
for T$_{\rm K}$=300~K. In this case the density is $\simeq$10$^{\rm 6}$~cm$^{\rm -3}$. In both cases the three lines are masing. The large dilution 
factors we derive imply a large number of small sources inside the beam. As a reference, the extremes of the estimated dilution factors of 0.0007 and 
0.005 would be equivalent to a single source of 0.018\arcsec\ (7~pc) and 0.05\arcsec\ (18~pc), or a large number of smaller sources. This is also what 
has been previously suggested by \citet{cer06} and \citet{gal16}.\\
\indent
The above estimations assume the same dilution factor applied for the three lines, which is of course very unlikely if we consider what is known for 
the emission of these lines in galactic sources (see Sect.\,\ref{subsec:line_morphology}), but it is the best we can do with the limited information 
we have so far on the spatial extent of the emission. Moreover, the lack of information about the position of the sources relative to the continuum 
dust emission introduces large uncertainties in the estimated densities and temperatures. In addition, the solution found fitting all three lines 
has to be taken with caution in view of the uncertainties in the emission of the 22~GHz line.\\
\indent
We conclude that the most plausible origin of the observed emission is that a large number of small dense and warm molecular clouds, strongly 
diluted in our 0.7\arcsec\ angular resolution synthetic beam, are present around the nuclei of Arp~220. A solution involving moderate densities of 
10$^{\rm 4-5}$~cm$^{\rm -3}$ and T$_{\rm K}$\,=\,100~K is also possible but fails to reproduce the 22~GHz line. For the same column density of 
water, the 183 and 325~GHz lines could be also fitted with higher densities if we assume a dilution factor for the emitting clouds. 
Due to the peculiar physical conditions needed to pump the 183 and 325~GHz lines (see Table\,\ref{tab:water_properties}) -- they originate from 
spatial sizes much larger than those corresponding to the 22~GHz maser emission \citep[e.g.,][]{cer90,cer94} -- higher spatial resolution observations 
with ALMA of these lines could provide strong constraints on the physical conditions of the molecular clouds of Arp~220: Assume that the ensemble of 
clouds from which the emission at 183 and 325~GHz originates in Arp~220 occupies similar size scales to e.g., the emission region in Sgr~B2 
\citep[about 25~pc, e.g.,][]{sco75}. Then the Arp~220 emission region would be at least a factor of 10 smaller than the beam size in our data 
($\sim$0.06\arcsec\ at the Arp~220 distance). Distinct regions exhibiting maser emission in Sgr~B2 have sizes of about 0.7~pc \citep[e.g.,][]{deV97}, 
which at the distance of Arp~220 correspond to source sizes of $\sim$0.002\arcsec. This would mean that we cover between 100 and $\sim$1000 Sgr~B2 
equivalent sources with our ALMA beam. Although we cannot resolve the individual sources, the longest baseline configurations of ALMA can help 
disentangle the distribution of water emission in Arp~220.

%______________________________________________________________

\section{Summary and conclusions} \label{subsec:summary_conclusions}

With our interferometric observations of three water lines (22~GHz, 183 and 325~GHz) we constrain the physical conditions in the H$_{\rm 2}$O 
emitting gas in both nuclei of Arp~220. A comparison of spectra of the 183 and 325~GHz water lines observed at different dates within a 
time frame of 11 and two years, respectively, shows that the emission is not variable given the velocity resolution and sensitivity of the data. 
The 22~GHz observations suggest that the lack of emission in the western nucleus at this frequency is most likely not intrinsic to the physics of 
the water line, but a result of the strong ammonia absorption. The observed line intensity ratios are not compatible with a pure thermal 
origin of the water emission. An LVG model does reproduce the observed results quite well for Arp~220 when introducing a dilution factor for the 
emitting clouds: column densities above 10$^{\rm 18}$~cm$^{\rm -3}$\,(km\,s$^{\rm -1}$)$^{\rm -1}$, temperatures T$_{\rm kin}$\,$\geq$\,200-300~K, 
and H$_{\rm 2}$ densities between $\sim$10$^{\rm 5}$ and 10$^{\rm 6}$~cm$^{\rm -3}$. Our findings support previous suggestions that a large 
number of star-forming clumps are the source of the bright water maser emission in both nuclei of Arp~220.

%-------------------------------------------------------------------

\begin{acknowledgements}

We thank the staff at the JAO and the EU ARC Network who have participated in the EOC and Science Verification activities, observations and 
data reduction that made the release of the data to the community possible. This paper makes use of the following ALMA data: 
ADS/JAO.ALMA\#2011.0.00018.SV and ADS/JAO.ALMA\#2012.1.00453.S. ALMA is a partnership of ESO (representing its member states), NSF (USA) and 
NINS (Japan), together with NRC (Canada) and NSC and ASIAA (Taiwan), and KASI (Republic of Korea), in cooperation with the Republic of Chile. The 
Joint ALMA Observatory is operated by ESO, AUI/NRAO and NAOJ. The National Radio Astronomy Observatory is a facility of the National Science 
Foundation operated under cooperative agreement by Associated Universities, Inc. IRAM is supported by INSU/CNRS (France), MPG (Germany) and IGN 
(Spain). JC and AF thank the ERC for support under grant ERC-2013-Syg-610256-NANOCOSMOS. They also thank Spanish MINECO for funding support under 
grants AYA2012-32032, and from the CONSOLIDER Ingenio program ``ASTROMOL'' CSD 2009-00038. KS acknowledges grant 105-2119-M-001-036 from the Taiwanese 
Ministry of Science and Technology.

\end{acknowledgements}

%-------------------------------------------------------------------

\bibliographystyle{aa}
\bibliography{arp220}

\begin{appendix}

\section{22~GHz correction} \label{sec:appendix_1}

\begin{figure*}[t]
  \centering
    \includegraphics[width=0.5\textwidth]{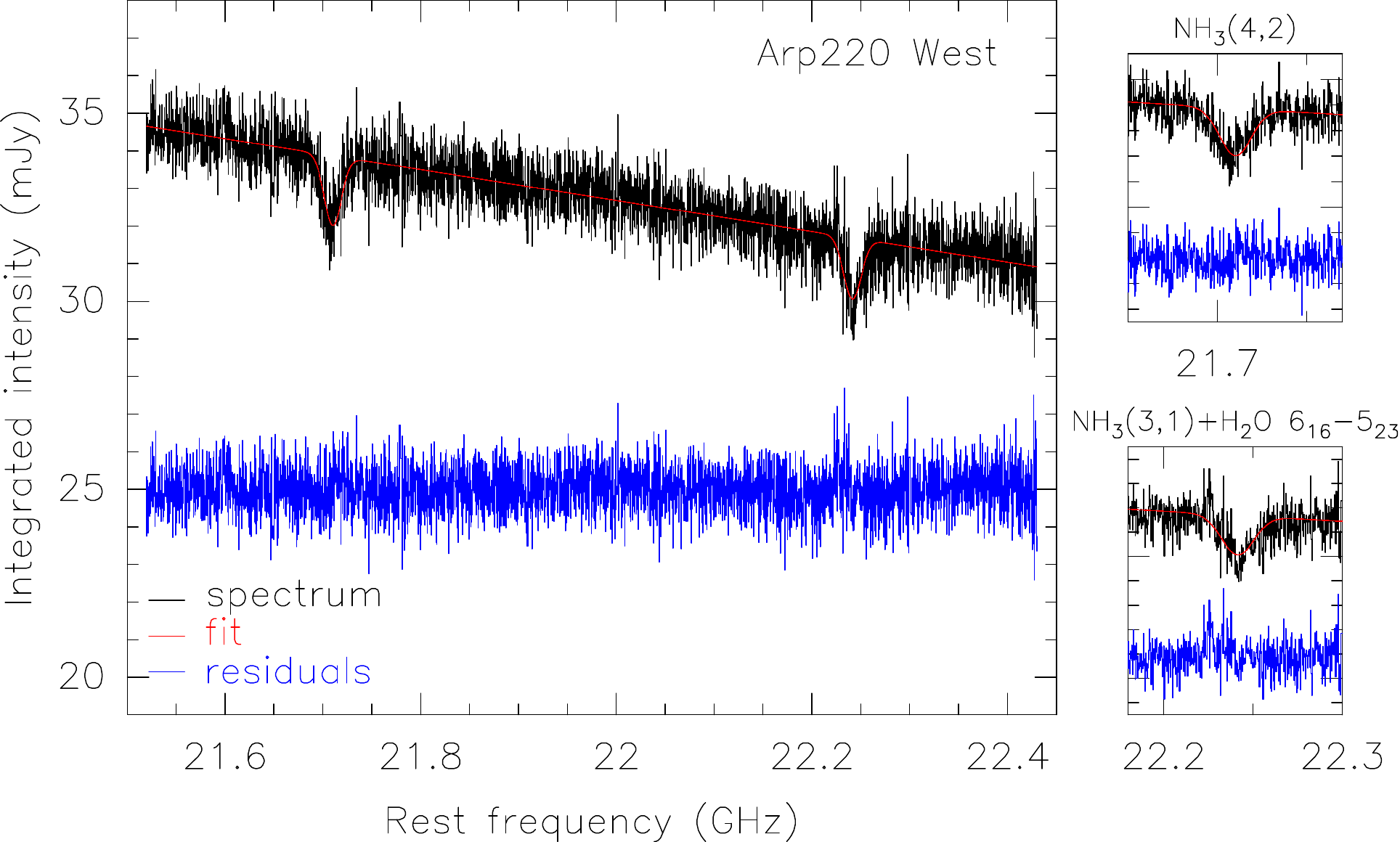}
  \caption{\footnotesize \textit{Left:} Image showing the original 22~GHz spectrum in Arp~220 West (in black), the applied fit to account for the 
   NH$_{\rm 3}$ absorption (red) and the residual spectrum (blue). For clarity purposes the residual spectrum is shown at an offset in y by 
   +25~mJy. \textit{Right:} Zoom into the NH$_{\rm 3}$\,(4,2) (\textit{top}) and NH$_{\rm 3}$\,(3,1) (\textit{bottom}) line features.}
  \label{fig:22ghz_comp}
\end{figure*}

\section{ALMA Band~5 spectrum} \label{sec:appendix_2}

\begin{figure*}[t]
  \centering
    \includegraphics[width=0.9\textwidth]{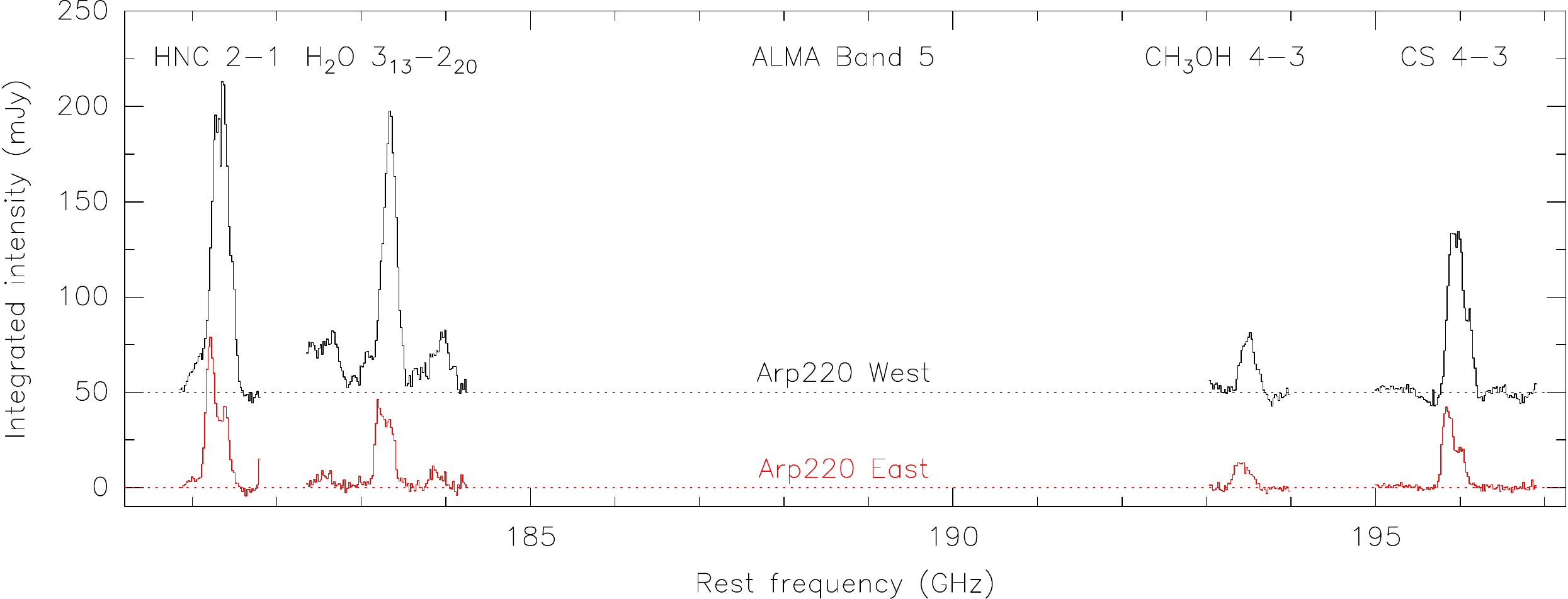}
  \caption{\footnotesize Spectrum covering all spectral windows observed with ALMA in Band~5. Arp~220 West spectra are artificially 
   offset by 50~mJy for clarity of the figure. The molecular transitions on which each spectral window was centered are indicated above.}
  \label{fig:spec_b5}
\end{figure*}

\begin{table*}[ht]%[!h]
\begin{minipage}[!h]{\textwidth}
\centering
\renewcommand{\footnoterule}{}
\caption{\small
 Properties of the additional emission lines securely identified in ALMA Band~5.}
\label{tab:properties_band_5_lines}
\tabcolsep0.1cm
\begin{tabular}{lccccccc}
\hline
\noalign{\smallskip}				% \newline \hspace*{3mm}
\hline
\noalign{\smallskip}
%---------------------------------------------------------------------------
Line & $\nu$$_{\rm rest}$  & \multicolumn{2}{c}{$\int$$S$\,d$v$\footnote{The left column denotes the integrated intensities measured in Arp~220 East, the right  column is for Arp~220 West.}\saveFN\inte}   & \multicolumn{2}{c}{Peak flux\useFN\inte} & \multicolumn{2}{c}{FWHM\useFN\inte}\\
\noalign{\smallskip}
     & [GHz]               & \multicolumn{2}{c}{[Jy\,km\,s$^{\rm -1}$]} & \multicolumn{2}{c}{[mJy]}     & \multicolumn{2}{c}{[km\,s$^{\rm -1}$]} \\
\noalign{\smallskip}
%---------------------------------------------------------------------------
\noalign{\smallskip}
\hline
\noalign{\smallskip}
HNC\,2$-$1             & 181.32476 & 36.4 & 91.4 & 79.0 & 163.0 & 469\,$\pm$\,20 & 369\,$\pm$\,20 \\
CH$_{\rm 3}$OH\,4$-$3  & 193.45436 & 8.1  & 12.8 & 13.0 & 31.3  & 376\,$\pm$\,20 & 346\,$\pm$\,20 \\
CS\,4$-$3              & 195.95421 & 18.5 & 35.8 & 42.2 & 84.5  & 437\,$\pm$\,20 & 437\,$\pm$\,20 \\
VIB-HC$_{\rm 3}$N\footnote{This line is a blending of different vibrational HC$_{\rm 3}$N components.} & $\sim$182.6 & 2.0 & 16.3 & 9.0 & 32.1 & --\footnote{Since this is a blending of several components, and the FWHM is not fully covered by the band, no values for the line width are given.}\saveFN\vib    & --\useFN\vib    \\
CH$_{\rm 3}$CN\,10$-$9 & 183.957   & 2.4  & 10.8 & 11.4 & 32.7  & 332\,$\pm$\,20 & 396\,$\pm$\,20 \\
\noalign{\smallskip}
\hline
%--------------------------------------------------------------------------
\end{tabular}
\end{minipage}
\end{table*}

\end{appendix}

\end{document}